# On the sensitivity, selectivity, sensory information and optimal size of resistive chemical sensors

(Invited paper)

Laszlo B. Kish [+], Janusz Smulko [++], Peter Heszler [*], and Claes-Goran Granqvist [**]

[+] *Department of Electrical and Computer Engineering, Texas A&M University, College Station, TX 77843-3128, USA*

[++] *Gdansk University of Technology, WETiI, ul. G. Narutowicza 11/12, 80-952 Gdansk, Poland*

[*] *Research Group of Laser Physics of the Hungarian Academy of Sciences, University of Szeged, P.O. Box 406, H-6701 Szeged, Hungary*

[**] *Department of Engineering Sciences, The Ångström Laboratory, Uppsala University, P.O. Box 534, SE-751 21 Uppsala, Sweden*

**Abstract.** Information theoretical tools are applied for the study of the sensitivity and selectivity enhancements of resistive fluctuation-enhanced sensors. General considerations are given for the upper limit of selectivity enhancement. The signal-to-noise ratio and information channel capacity of fluctuation-enhanced chemical sensors is compared to that of classical sensors providing a single output. The considerations are done at the generic level with a few concrete examples and include the estimation of scaling relations between the sensor size, speed of measurements, sampling rate, measurement time, signal power and noise power.

**1. Introduction.**

*1.1 Classical gas sensing*

*Gas sensors for healthy environments*
Concerns about *outdoor* air-pollution are widely spread. However, it is less known that serious health-related problems may emerge from the *indoor* environment, too. Indoor air contains a wide variety of volatile organic compounds (VOCs, e.g. formaldehyde, vapors of organic solvents), and a number of these VOCs have a higher concentration indoors than outdoors [1]. Exposure to VOCs has been suggested to cause e.g., mucous irritation, neurotoxic effects (fatigue, lethargy, headache etc.) and nonspecific reactions (e.g. chest sounds and asthma-like symptoms) [2, 3]. It is clear that precise air quality monitoring is of great importance in both in- and outdoor environments. This requires sensors capable of detecting low concentrations of $CO_2$, CO, $SO_2$, $NO_x$, $O_3$, $H_2S$, HF, $Cl_2$, VOCs, etc. sensitively and selectively. (The listed gases have been selected, as they have toxic effects [4].) This huge need could be best fulfilled with simple, cheap and replaceable sensors, most preferably electronic, semiconductor type, that can be easily integrated into the existing monitoring and ventilation systems.

*Semiconductor gas sensors*

The operation principle of the "classical", Taguchi-type of semiconductor gas sensors is based on the change of the sensor resistance as the gas to be sensed is adsorbed on the sensor surface [5]. This type of sensors represents a low-cost option to the standardized and bulky methods (e.g., gas chromatography or mass spectroscopy). Mostly metal-oxides, e.g. $SnO_2$, $TiO_2$, ZnO, $Mn_2O_3$ and $WO_3$, are used as sensor materials [6]. There is a continuous work for improving the sensor performance including sensitivity and most importantly, the chemical selectivity of these kinds of sensors.

Toward sensitivity, nanotechnology, more closely the use of Nanostructured Materials (NsM), offers new possibilities in this area, too. In general, the characteristic structural length of a NsM is typically 1 to 100 nanometers. One class of NsMs is composed of nanoparticles or nanocrystals, and in a porous structure these materials exhibit high surface area, which can be orders of magnitude higher than that of coarser, micro-grained materials, therefore increasing sensitivity of the gas sensors [7, 8]. Likely, not only the high surface area, but the actual nanostructure (e.g., neck and grain boundary formation between nano-grains) also plays role of sensitivity improving of NsMs [9]. Sensitivity can also be improved by doping the oxide materials [6, 10].

Chemical selectivity of semiconductor gas sensors can be improved by operating an array of sensors, each of them having different sensitivity for different gases (can also be called electronic nose) [11]. This can be achieved by e.g., using different (or doped) sensor materials or by operating the sensors at different temperatures. The output of sensor arrays is then analyzed by pattern recognition methods [12]. Analyzing the dynamic response of temperature-modulated sensors is also a possible way for improving chemical selectivity [13]. However, lack of selectivity is still a significant problem for the widespread use of semiconductor gas sensors.

*1.2 Fluctuation-enhanced sensing*

While some optical chemical sensors analyze the absorption or emission spectrum of gases and therefore able to generate a pattern, most of chemical sensors produce a single number output only. For example the steady-state value of a Taguchi sensor or the steady-state current value of a MOS sensor are such signals. To generate a separate pattern corresponding to different chemical compositions, a number (6-40) of different types of sensors are needed which makes the system expensive and unreliable for practical applications. On the other hand Fluctuation-Enhanced Sensing (FES) is able to generate a complex pattern by the application of a single sensor [14-22]. FES means that, instead of using the mean value (time average) of the sensor signal, the small stochastic fluctuations around the mean value are amplified and statistically analyzed. Due to the grainy structure of resistive film sensors, these materials exhibit significantly (several orders of magnitude) increased electronic resistance fluctuations compared to single crystalline materials and these fluctuations are strongly influenced by the random walk (diffusion) dynamics of agents in the vicinity of intergrain junctions and by adsorption-desorption noise. Stochastic analytical tools are used to generate a one-dimensional of two-dimensional pattern from the time fluctuations. The analysis of these patterns can be done in the classical way by using pattern recognition tools.

The history of FES is more than a decade long [14-37]. The name "Fluctuation-Enhanced Sensing" was created by John Audia (SPAWAR, US Navy) in 2001. Here we mostly focus on journal papers and neglect the vast body of conference contributions except in cases where patents or conference papers have given the priority.

Using the electrical noise (spontaneous fluctuations) to identify chemicals was first proposed by Bruno Neri and coworkers [14,15] in 1994-95 by showing the sensitivity of conductance noise spectra of conducting polymers against the ambient gas composition. In 1997, Gottwald and coworkers [16] published similar effects with the conductance noise spectrum of semiconductor resistors with non-passivated surface. The first mathematical analysis of generic FES systems with the sensor number requirement versus the number of agents was done by Kish and coworkers in 1998 [17-19]. The possibility of "freezing the smell" in the Taguchi sensor was first demonstrated by Robert Vajtai [18] and later a more extensive analysis published by Solis et al [20]. In 2001, Smulko et al have first time used Higher-Order Statistics (HOS) to enhance the extracted information from the stochastic signal component [21,26,29]. Hoel et al showed FES via invasion noise effects in room-temperature nanoparticle films [22]. Schmera and coworkers analyzed the situation of Surface Acoustic Wave (SAW) sensors and predicted the FES spectrum for SAW and MOS sensors with surface diffusion [23-24]. Commercial-On-The-Shelf (COTS) sensors with environmental pollutants and gas combinations were also studied [25,29,30]. In nanoparticle sensors with temperature gradient, the possibility of using the noise of the thermoelectric voltage for FES was demonstrated [28]. Ederth et al analyzed and compared the sensitivity enhancement in the FES mode compared to the classical mode in nanoparticle sensors and found an enhancement of a factor of 300. Gomri et al [32,33] published FES theories for the cases of adsorption-desorption noise and chemisorption-induced noise. Huang et al explored the possibility of using FES in electronic tongues [34].

## 2. Sensitivity and selectivity enhancement in fluctuation-enhanced sensing

The statistics of the microscopic fluctuations in a system are rich and sensitive sources of information about the system itself. They are extremely sensitive because the perturbations of microscopic fluctuations require only a very small energy. On the other hand, the related statistical distribution functions are data arrays thus they can contain orders of magnitude more information then a single number represented by the mean value of sensor signal used in classical sensing.

The underlying physical mechanism behind the enhanced *sensitivity* are the temporal fluctuations of the agent's or its fragment's concentration at the various points of the sensor volume where the sensitivity of the resistivity against the agent is different. This effect will generate stochastic fluctuations of the resistance and the sensor voltage during biasing the sensor with a DC current. The voltage fluctuations can be extracted (by removing the mean value by AC coupling) and strongly amplified. The significantly increased sensitivity was demonstrated in several experiments, see and sensitivity enhancement by several orders of magnitude has been demonstrated by Kish and coworkers [29] in Taguchi sensors and Ederth and coworkers [31] in nanoparticle films.

Significantly increased *selectivity* can be expected depending on the type of sensor and types of available FES fingerprints. We define the selectivity enhancement by the factor of how many classical sensors a fluctuation-enhanced sensor can replace. When using power density spectra, the theoretical upper limit of selectivity enhancement is equal to the number of spectral lines. At typical experiments that is about 10000. However, when the elementary fluctuations are random-telegraph signals (RTS) the underlying elementary spectra are Lorentzians [35,36]and the situation is less favorable because their spectra strongly overlap. As a consequence, experiments with COTS sensors indicate that the response of spectral lines against agent variations is often not independent. In simple experimental demonstration with COTS sensors a selectivity enhancement of 6 was easily reachable [18]. However, nano sensor development may be able to use all the spectral lines more independently. Because both the FES signal in macroscopic sensors and the natural conductance fluctuations of the resistive sensors usually show 1/f like spectra [35,36], the lower the inherent 1/f noise strength in the sensor the cleaner the sensory signal. An interesting analysis can me made, if we suppose we shrink the sensor size so much that the different agents probe different RTS signals. Then 1/f noise generation principles [37,38] indicate that one can resolve at most a few Lorentzian components in a frequency decade and supposing 6 decades of frequency, the maximal selectivity enhancement would be around 18, supposing 3 fluctuators/decade.

With bispectra [21,26,27], the potential of selectivity-increase is even greater because bispectra are two-dimensional data arrays. In the case of 10000 spectral lines mentioned above, the theoretical upper limit of selectivity increase is 100 million, however, in the Lorentzian fluctuator limit, that number is also radically reduced. Because bispectra sense only the non-Gaussian part of the sensor signal, for the utilization of the full advantages of bispectra, it seems it is necessary to build the sensor for the submicron characteristic size range to utilize elementary microscopic switching events as non-Gaussian components. Moreover, the sixfold symmetry of the bispectrum function yields a further reduction of information by roughly a factor of 6. Using the above-mentioned estimation with 3 Lorentzian fluctuators/decade, over 6 decades of frequency the selectivity enhancement would be around 50. Note that this enhancement is independent from the spectral enhancement discussed above because bispectra probe the non-Gaussian components,

## 3. Signal to noise ratio and information channel capacity with classical sensors

Claude Shannon was following Nyquist's [39] and Leo Szilard's [40] pioneering breakthroughs of using the entropy and the bit as the measure of information and for white noise and signal spectra he found [41] that the information channel capacity , which is the upper limit of possible information flow rate given in *bit/s* is:

$$C = W \ln\left(1 + \frac{P_S}{P_N}\right) \quad (1)$$

where, in the memory-less limit, $W$ is the bandwidth, $P_S$ is the mean-square signal voltage (signal "power") and $P_N$ is the mean-square noise voltage (noise "power"). This equation can be rewritten for measurement time duration $t_m$ by using Shannon's sampling theorem:

$$C = \frac{1}{2t_m} \ln\left(1 + \frac{P_S}{P_N}\right) \quad (2)$$

If the sensor resistance is measured by a constant current generator driving then, in accordance with Ohm's law, the signal "power" is:

$$P_S = I^2(R - R_0)^2 \, , \quad (3)$$

where $I$ is the current, $R$ is the resistance in the agent (test) gas and $R_0$ is the resistance in the reference gas. The noise in a resistive sensor with macroscopically homogeneous current density (when contact noise is neglected) is the superposition of thermal noise and 1/f-like noises [14-21]:

$$S_{u,N}(f) = 4kTR + AR^2I^2V^{-1}f^{-\gamma} \qquad (4)$$

where $S_{u,N}(f)$ is the power density spectrum of the noise voltage on the sensor, $k$ is the Boltzmann constant, $T$ is the absolute temperature, $V$ is the volume of sensor film, and $A$ is the 1/f noise coefficient of the material (normalized 1/f noise spectrum in unit volume) [35, 36] and $\gamma$ is the frequency exponent (~ 0.8 - 1.3).

The noise "power" can be determined by the well-known relation:

$$P_N = \int_{f_1}^{f_2} S_{u,N}(f)df \qquad (5)$$

where, in the thermal noise dominated limit the $f_1 = 0$ and $f_2 \approx 1/t_m$ approach holds. When the 1/f noise is the dominant already around frequencies $f_2 \approx 1/t_m$, then we are in the constant noise "power" limit due to the constant variance of time-averaged 1/f noise:

$$P_N \approx \frac{AR^2I^2}{8\pi^2 V} \qquad (6)$$

This limit is the practical one because of the strong 1/f noise and the relatively long time (ranging from millisecond to several minutes) resistive sensors need to produce a stationary resistance in a changed ambient gas.

In conclusion, classical resistive sensors have the following upper limit of information flow rate:

$$C = \frac{1}{2t_m} \ln\left[1 + \frac{8\pi^2 V(R-R_0)^2}{AR^2}\right] = \frac{1}{2t_m} \ln\left[1 + \frac{8\pi^2 A_S d(R-R_0)^2}{AR^2}\right] \qquad (7)$$

where $A_S$ is the surface of the sensor film and $d$ is its thickness. According to Eq. (7), in the practical (1/f noise dominated) limit, at fixed measurement time and film thickness, the larger the surface of the classical resistive sensor the greater the information channel capacity. However, in the sufficiently large agent concentration limit, the saturation time is controlled by the underlying diffusion processes through the thickness $d$ of the film, therefore, in this case, the shortest measurement time in the is also controlled by diffusion:

$$t_{m,min} \approx \left(\frac{d}{D}\right)^2 \qquad (8)$$

where $D$ is the diffusion coefficient of the agent and/or its fragments through the film. Therefore, the thinner the film the faster the response and greater the information channel capacity. This fact indicates that in classical films small thickness and large surface is preferable.

## 4. Information channel capacity at fluctuation-enhanced sensing

In the case of FES, the signal is the change of statistical parameters of the measured FES voltage while the sensor is exposed to the agent compared to the situation while the sensor is exposed to the reference gas, such as synthetic air.

### 4.1 Power density spectrum based sensing

According the Shannon [41], when the signal and/or the noise have colored spectrum, then the following relation is in effect:

$$C = \int_0^B \ln\left(1 + \frac{S_S(f)}{S_N(f)}\right) df \quad , \tag{9}$$

where $B$ is the upper cutoff frequency, $S_S(f)$ and $S_N(f)$ are the signal and the noise spectrum, respectively. Because the FES power density spectrum is colored in both cases (while the sensor is exposed to the agent and while it is exposed to the reference gas) it may look tempting to use Eq. (9) with the agent generated spectrum as "signal spectrum" and reference gas generated spectrum as "noise spectrum". However such a use would be incorrect for several reasons. For example, according to Solis, et al [30] the spectra generated by different agents are not additive in commercial Taguchi sensors which is probably due to the nonlinear mixing of the noise dynamics at the elementary fluctuator level during the exposure to the diffusion processes of different fragments in the vicinity of the intergrain junctions. Another reason is that in reality, the FES signal is the *change* of the power spectrum $S(f)$ of the measured FES voltage while the sensor is exposed to the agent compared to the reference spectrum $S_0(f)$ while the sensor is exposed to the reference gas, such as synthetic air. Note, the spectrum in the reference gas is not background noise but it is also a signal itself. The background noise is related to the statistical inaccuracy of the measured spectra due to finite-size/finite-time statistics. If we suppose that the statistical inaccuracies with the test gas and the reference gas are independent and they have then:

$$C = \frac{1}{t_m} \sum_{i=1}^{M} \frac{1}{2} \ln\left\{1 + \frac{[S(i\Delta f) - S_0(i\Delta f)]^2}{\Delta S^2(i\Delta f) + \Delta S_0^2(i\Delta f)}\right\} \tag{10}$$

where $M$ is the number of separate frequency bands (supposed to have uniform bandwidth $\Delta f$) having independent spectra. The *rms* error of the average spectrum in one frequency band can be estimated as:

$$\frac{\Delta S(f)}{S(f)} \approx \frac{1}{K\sqrt{t_w \Delta f}} = \frac{1}{K\sqrt{N f_s^{-1} \Delta f}} = \frac{1}{\sqrt{K t_m \Delta f}} \quad , \tag{11}$$

where $K$ is the number of data sequences used for the determination of the average spectrum, $t_w$ is the duration (time window) of a single data sequence, $N$ is the number of data in a single data sequence and $f_s$ is the sampling frequency.

supposing equal frequency bands and supposing that the relative error of the spectrum is much less than 1, thus the logarithmic term can be approximated with a constant:

$$C \propto t_m^{-1} M = \frac{t_m t_w f_s^2}{\Delta f} \tag{12}$$

where we used the following relations:

$$t_m = \frac{KN}{f_s} \tag{13}$$

$$M = \frac{t_w}{t_s} \frac{f_s}{\Delta f} = \frac{t_w f_s^2}{\Delta f} \tag{14}$$

### 4.2 Amplitude-distribution-based sensing

A similar equation can be deduced for the amplitude distribution method:

$$C = \frac{1}{t_m} \sum_{i=1}^{M} \frac{1}{2} \ln\left\{1 + \frac{[g(i\Delta U) - g_0(i\Delta U)]^2}{\Delta g^2(i\Delta U) + \Delta S_0^2(i\Delta U)}\right\} \tag{15}$$

where $M$ is the number of separate amplitude bands (supposed to have uniform bandwidth $\Delta U$) having independent statistics. In the present paper, we are focusing on the spectral methods however relations with similar nature as in Section 3 can be deduced. However, it is important to note that, similarly to bispectra, the only meaningful case of FES based on amplitude distributionis the situation where the amplitude distribution is non-Gaussian.

### 4.3 Bispectrum-based sensing

The same principles, as applied above for power spectra, for the case of bispectrum $B$ lead to the following information channel capacity:

$$C \approx \frac{1}{6} \frac{1}{t_m} \sum_{i=1}^{M} \sum_{j=1}^{M} \frac{1}{2} \ln\left\{1 + \frac{[B(i\Delta f, j\Delta f) - B_0(i\Delta f, j\Delta f)]^2}{\Delta B^2(i\Delta f, j\Delta f) + \Delta B_0^2(i\Delta f, j\Delta f)}\right\} \qquad (16)$$

where the variance $\Delta B$ of bispectrum estimation represents the background noise amplitude and the term 1/6 is the estimation of the reduction due to the sixfold symmetry of the bispectrum.

In the paper of Nikias and Mendel [42], the variance $\Delta B^2(f_1, f_2)$ of bispectrum estimation is given for rectangular time window as:

$$\Delta B^2(f_1, f_2) = \frac{N}{K} S(f_1) S(f_2) S(f_1 + f_2), \qquad (17)$$

where $f$ is the frequency, $N$ is the number of data points in a single data sequence. $K$ is the number of data sequences, and $S(f)$ is the power density spectrum of the analyzed signal. Supposing a single signal (no ensemble averaging), the total number of data point is $L = KN$ and the total length of measurement is $t_m = L/f_s$. The accessible band of frequency is $f_{low} = f_s/N < f < f_{high} = f_s/2$, where $f_{low}$ and $f_{high}$ are the low-frequency and the high-frequency limits, respectively. Thus, Eq (17) can be written in a more practical form:

$$\Delta B^2(f_1, f_2) = \Delta B^2(f_1, f_2) = \frac{f_s t_m}{K^2} S(f_1) S(f_2) S(f_1 + f_2) = \frac{N^2}{f_s t_m} S(f_1) S(f_2) S(f_1 + f_2) \qquad (18)$$

The $\Delta B^2(f_1, f_2)$ value determined by Eq. (18) then serves as the input value for the bispectrum inaccuracies $\Delta B^2(i\Delta f, j\Delta f)$ and $\Delta B_0^2(i\Delta f, j\Delta f)$ in Eq. (16).

## 4. Conclusion

A Taguchi sensors (resistive grainy film sensors) should be thin for fast response (short $t_m$) and therefore for large information channel capacity. Moreover, for classical sensing, the greater the size and smaller their 1/f noise factor (*in the presence of agents!*) the greater their sensory information channel capacity. In the case of FES, the smaller the size is the more sensory information up to the point that the signal becomes strongly non-Gaussian, so amplitude distribution function a bispectra can also be utilized. The information channel capacity will also be influenced by the choice of the single measurement time window $t_w$. The optimal choice depends also on the characteristics of pattern recognition technique applied to identify the agent composition.